\documentclass[floatfix, twoside, twocolumn, aps]{revtex4}
\usepackage{epsfig}
\begin{document}

\title{Order book approach to price impact}

\author{Philipp Weber and Bernd Rosenow}

\affiliation{Institut f\"ur Theoretische Physik, Universit\"at zu
K\"oln, D-50923 Germany}

\date{December 9, 2003}

\begin{abstract}

Buying and selling stocks causes price changes, which are
described by the price impact function. To explain the shape of
this function, we study the Island ECN orderbook. In addition to
transaction data, the orderbook contains information about
potential supply and demand for a stock.  The virtual price impact
calculated from this information is four times stronger than the
actual one and explains it only partially. However, we find a
strong anticorrelation between price changes and order flow, which
strongly reduces the virtual price impact and provides for a
quantitative explanation of the empirical price impact function.

\end{abstract}

\maketitle

In a perfectly efficient market, stock prices change due to the
arrival of new information about the underlying company.  From a
mechanistic point of view, stock prices change if there is an
imbalance between buy and sell orders for a stock.  These ideas can be
linked by assuming that someone who trades a large number of stocks
might have private information about the underlying company, and that
an imbalance between supply and demand transmits this information to
the market.  In this sense, order imbalance and stock price changes
should be connected causally, i.e.  prices go up if demand exceeds
supply and go down if supply exceeds demand.  The analysis of huge
financial data sets \cite{Takayasu02} allows a detailed study of the
price impact function
\cite{Hasbrouck91,HaLoMc92,kempf99,pler2002,Rosenow02,EvLy02,LiFaMa03,Ga+03,potbou2002,Hop02,Bou+03},
which quantifies the relation between order imbalance and price
changes.

Potential supply and demand for a stock is stored in the limit
order book. If a trader is willing to sell a certain volume
(number of shares) of a stock at a given or higher price, she
places a limit sell order. For buying at a given or lower price, a
limit buy order is placed. An impatient trader  who wants to buy
immediately places a market buy order, which is matched with the
limit sell orders offering the stock for the lowest price, the ask
price $S_{\rm ask}$ for that stock.  Similarly, a market sell
order is matched with the limit buy orders offering the highest
price, the bid price $S_{\rm bid}$.

In previous studies
\cite{Takayasu02,Hasbrouck91,HaLoMc92,kempf99,pler2002,Rosenow02,EvLy02,LiFaMa03,Ga+03,potbou2002,Bou+03}
(with the exception of \cite{Hop02}), the price impact of trades
was calculated by determining whether a given trade was buyer or
seller initiated \cite{LeRe91}. Here, we analyze order book data
which unambiguously allow to identify the character of a
transaction. We first calculate the price impact of market orders,
which are aggregated in time intervals of length $\Delta t = 5
{\rm min}$, and compare it to the virtual or instantaneous price
impact, which would be caused by a market order matched with limit
orders from the order book.  The virtual price impact is found to
be four times stronger than the actual one.  To explain this
surprising discrepancy, we study time dependent correlations
between order flow and returns and find that limit orders are
anticorrelated with returns in contrast to the positive
correlations between returns and market orders.  We suggest that
limit orders placed in response to returns provide for a
quantitative link between virtual and actual price impact.

We analyzed data from the Island ECN, NASDAQ's largest electronic
communication network, which comprises about 20 percent of all trades.
We chose the 10 most frequently traded stocks for the year 2002
\cite{ticker}. The volume of market buy orders is counted as positive
and the volume of market sell orders as negative, and the sum of all
signed market orders placed in the time interval $[t,t+\Delta t]$ with
$\Delta t = 5 {\rm min}$ is denoted by $Q(t)$ \cite{hidden}. Stock
price changes are measured by the return $G(t)$ in the same time
interval as
%
%******************* return definition ******************************
\begin{equation}
G(t) = \ln S_M(t+\Delta t) - \ln S_M(t),
\end{equation}
%********************************************************************
%
where the midquote price $S_M(t)={1 \over 2} (S_{\rm bid}(t) + S_{\rm
  ask}(t))$ is the arithmetic mean of bid and ask price.  To make
different stocks comparable, we normalize the return time series
$G$ by their standard deviation $\sigma_G$
%= \sqrt{\langle G^2(t)\rangle - \langle G(t) \rangle ^2}$,
and the volume time series $Q$ by $\sigma_Q=\left \langle | Q -
\langle Q
  \rangle| \right \rangle$ as their second moment is not well defined
due to a slow decay of the probability distribution.\\

\begin{figure}
\centerline{ \epsfig{file=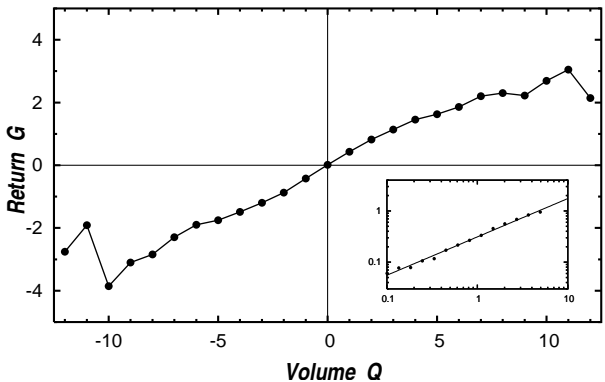,width=8cm,height=5cm}}
\caption{The price impact function
$I_{\rm market}(Q)$ for market
  orders is a monotonously increasing and concave function of the
  signed market order volume. A logarithmic plot (inset) shows that
  the function can be fitted by a power law.}
\label{priceimpact.fig}
\end{figure}

{\bf Price impact of market orders:} We define the price impact of
market orders as the conditional expectation value
%
%*********************** price impact definition ******************
\begin{equation}
I_{\rm market}(Q) = \langle G_{\Delta t}(t) \rangle_{Q}
\end{equation}
%*****************************************************************
%
for overlapping time intervals of market order flow and returns.
The functional form of $I_{\rm market}(Q)$ is shown in
Fig.~\ref{priceimpact.fig}. We find that $I_{\rm
  market}(Q)$ is a concave function of volume \cite{Hasbrouck91},
which can be well fitted by a power law $G = 0.48 \; Q^{0.76}$ with
$R^2 = 0.997$.  We note that the exponent $0.76$ as compared to $0.5$
found in \cite{Zhang99,pler2002,Ga+03} is due to the fact that we
compute returns for midquote prices as compared to returns for
transaction prices.  The concave shape of the function is very
surprising: This type of price impact would theoretically be an
incentive to make large trades as they would be less costly. In
contrast, a convex price impact would encourage a trader to brake up a
large trade into several smaller ones, which is what actually happens.
Having this in mind, we want to understand the mechanism responsible
for this concave shape and
analyze the trading information contained in the limit order book.\\

%
%**************************** figure (virtual) price impact *************
%***********************************************************************

{\bf Order book and virtual price impact:} At each instant in time
and for each stock $i$, the limit order book can be described by a
density function $\rho_i(\gamma,t)$ for the number of limit orders
where
%
%****************************  definition of return ********************
\begin{equation}
\gamma = \left\{ \begin{array}{ccc}
(\ln(S_{\rm{limit}}) - \ln(S_{\rm{bid}})) & \mbox{limit buy order}\\
(\ln(S_{\rm{limit}}) - \ln(S_{\rm{ask}})) & \mbox{limit sell
order}
\end{array}\right.   \  .
\label{return.eq}
\end{equation}
%*************************************************************************
%
We reconstructed the time dependent density functions for all ten
stocks from information about placement, cancellation, and
execution of limit orders contained in the Island ECN data base,
thereby processing about 60GB of data.

\begin{figure}
  \centerline{ \epsfig{file=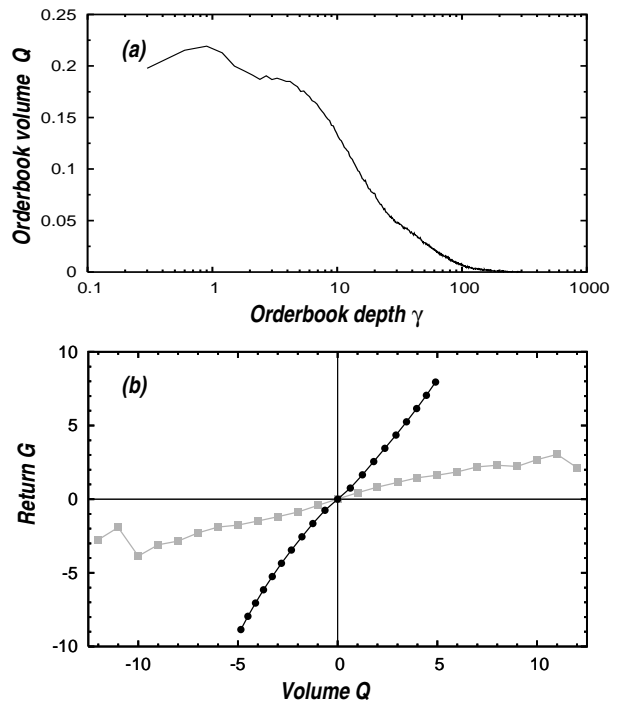,width=8cm} } \caption{(a) The
  average order book is characterized by a maximum at $\gamma = 1$ and
  a slow decay up to $\gamma = 100$. The negative side (buy orders) of
  the order book is similar to the positive one. (b) The virtual price
  impact function $I_{\rm book}(Q)$ (circles) calculated from the
  average limit order book is a convex function of order volume and
  much steeper than the price impact of market orders (squares).}
\label{flowcomp.fig}
\end{figure}

First, we study the average order book $\rho_{\rm{book}}(\gamma) =
\langle \rho_{i}(\gamma,t) \rangle$, where $\langle ... \rangle$
denotes an average over both time and different stocks. It is
characterized by a flat maximum at $\gamma \approx 1$ and a slow
decay for large $\gamma$ (Fig.~\ref{flowcomp.fig}a). Its overall
shape agrees with the results of \cite{MaMi01,ChSt01,Bouchaud02}.
We note that we have approximated $\rho_{\rm{book}}(G)$ on a grid
with spacing $0.3 \; \sigma_G$.

%
%**************************  figure mean order flow **********************
%********************************************************************
%

Consider a trader who wants to buy a  volume $Q$ of
stocks and has only offers from the order book available.
Beginning at the ask price, she executes as many limit orders as
necessary to match her market order, and changes the ask price by
an amount of $G$. Traded volume $Q$ and return $G$ are related by
%
%************************ inverse price impact function ******************
\begin{equation}
Q_{\rm{book}} = \int_0^{G} \rho_{\rm{book}}(\gamma) \; d\gamma \ \
. \label{inversepif.eq}
\end{equation}
%*************************************************************************
%
%
The virtual price impact $I_{\rm book} (Q)$ is obtained by
inverting this relation \cite{average}.  We assume that the
bid-ask spread remains constant in the process and that the
midquote price changes by the same amount as the ask price. The
virtual price impact is four times stronger than the price impact
of actual market orders (see Fig.~\ref{flowcomp.fig}b), a volume
of $5 \sigma_Q$ causes a virtual price change of $8 \sigma_G$ but
only an actual price change of $2 \sigma_G$.  In addition, $I_{\rm
book} (Q)$ is a convex function that can be fitted well by a power
law $G = 1.22 \; Q_{\rm book}^{1.19}$ with an $R^2 = 0.998$ and
not a concave function as $I_{\rm
  market}(Q)$.  The average order book and thus the virtual price
impact can be decsribed by ``zero intelligence models''
\cite{Bouchaud02,Farmer}, in which orders are placed randomly. As the
virtual price impact is not a good approximation for the actual one,
it seems that an additional mechanism describing ``intelligent'' or
collective behavior is needed to explain it.
\\

%
%*********************** figure correlation function *****************
\begin{figure}
\centerline{ \epsfig{file=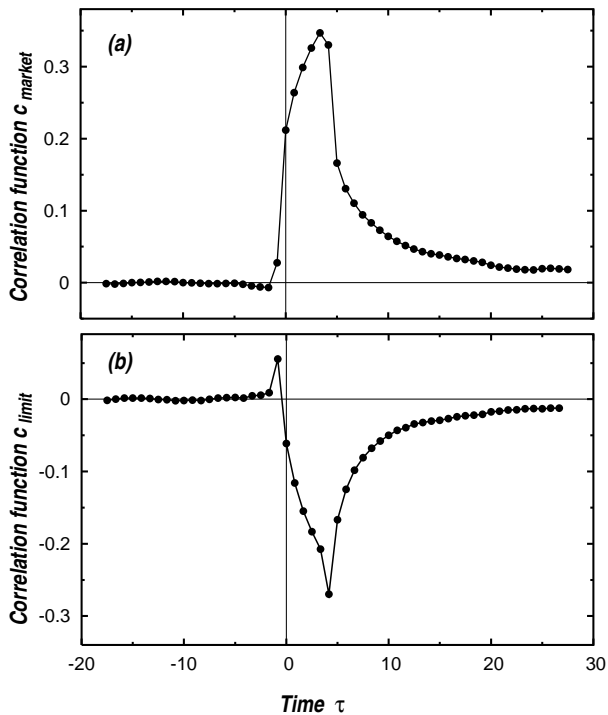,width=8cm} }
\caption{Correlation functions
between return and signed order flow (buy minus sell orders). (a)
Market orders and returns show strong positive equal time
correlations decaying slowly to zero.  (b) Limit orders preceding
returns have weak positive correlations with them, while equal
time correlations are strongly negative.}
\label{corrfunc.fig}
\end{figure}
%******************************************************************

{\bf Correlations between order flow and returns:} Which effect is
responsible for the pronounced difference between virtual and actual
price impact? In the following, we will argue that a strong
anticorrelation between returns and limit orders reduces the virtual
price impact and provides for the link between virtual and actual
price impact.  In order to understand how order flow and price changes
are related, we study the correlation functions
%
%******************** price--order correlation ********************
\begin{eqnarray}\hspace*{-.5cm}
c_\alpha(\tau)& =& {\langle  Q_\alpha(t+\tau) G(t)\rangle -
\langle Q_\alpha(t)\rangle \langle G(t) \rangle \over
\sigma_{Q_\alpha} \sigma_G}
\end{eqnarray}
%*****************************************************************
%
between the volume of market orders ($\alpha=\rm{market})$ or limit
orders ($ \alpha=\rm{limit}$) and returns.  The order volume is
measured in intervals $[t, t + \delta t]$ with width $\delta t= 50s$,
and the returns are recorded for five minute intervals.  For $\alpha =
{\rm market}$, $Q_{\rm market}(t)$ is the volume of signed market
orders, and for $\alpha = {\rm limit}$
%
%***************************** integrated flow of limit orders ********
\begin{equation}
Q_{\rm limit}(t) = \int_{-\infty}^{\infty}\!\!\!\! {\rm
sign}(\gamma) \;(Q^{\rm add}_{\delta t}(\gamma) - Q^{\rm
canc}_{\delta t}(\gamma)) \;d \gamma
\label{integratedlimitflow.eq}
\end{equation}
%*********************************************************************
%
is the net volume of limit sell orders minus the net volume of limit
buy orders. In Eq.~\ref{integratedlimitflow.eq}, $Q^{\rm add}_{\delta
  t}(\gamma)$ is the volume of limit orders added to the book at a
depth $\gamma$, and $Q^{\rm canc}_{\delta t}(\gamma)$ is the volume of
orders canceled from the book.

The correlation functions are plotted in Fig.~\ref{corrfunc.fig}. We
find that $c_{\rm market}(\tau)$ is zero for $\tau < -50s$ as required
for an efficient market where returns cannot be predicted over
extended periods of time. For times $\tau \ge -50s$, we find positive
correlations which are strongest when the time intervals for orders
and returns overlap. For $\tau > 250s$ (non overlapping time
intervals), we observe a slow decay of the correlation function which
is probably caused by the strong autocorrelations of the market order
flow \cite{Hasbrouck91,Hop02,LiFa03}.

The correlation function between limit orders and returns vanishes for
negative times $\tau < - 50 s$ and has a small positive value $c_{\rm
limit}(-50s)= 0.04$.  Surprisingly, for zero and positive time
differences there is a significant anticorrelation between limit
orders and returns, which is strongest for $\tau = 250s$ (overlapping
time intervals) and decays slowly to zero for large positive times. We
interpret this anticorrelation as an indication that rising prices
cause an increased number of sell limit orders and vice versa for
falling prices. Price changes seem to be counteracted by an
orchestrated flow of limit orders.
\\
%
%********** figure generated order flow and theoretical price impact ********
\begin{figure}
\centerline{ \epsfig{file=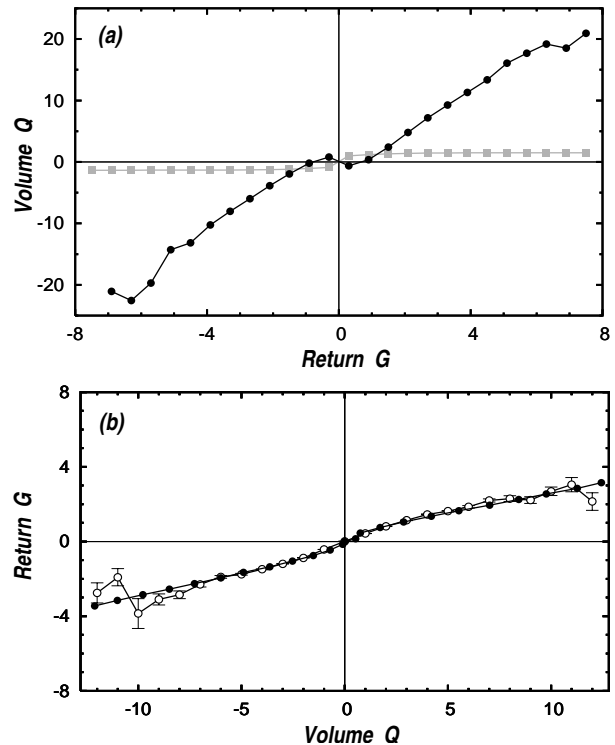,width=8cm} }
\caption{(a) The average flow of limit orders integrated up to an order book depth
$G$ (squares) changes rapidly at small returns and stays constant
then.  The additional volume of limit orders $Q_{\rm corr}$ in
response to a return $G$ (circles) increases linearly for large $G$.
(b) Empirical price impact $I_{\rm market}(Q)$ of market orders (open
circles) compared to the theoretical price impact $I_{\rm theory}(Q)$
(full circles), which takes into account orders from the order book,
the average flow of limit orders and the additional flow $Q_{\rm
corr}$.}
\label{gen_flow.fig}
\end{figure}
%*****************************************************************
%

{\bf Limit order flow and  feedback mechanism:} The
anticorrelation between returns and the flow of limit orders
suggests that dynamical effects are responsible for the difference
between virtual and actual price impact.  First, one should take
into account the average order flow.  The average density of this
order flow is described by
%
%************************* order flow  *********************************
\begin{equation}
\rho_{\rm flow}(\gamma) = \langle Q^{\rm add}_{\Delta t}(\gamma) -
Q^{\rm canc}_{\Delta t}(\gamma) \rangle \label{flow.eq}
\end{equation}
%*************************************************************************
%
with $\Delta t = 5{\rm min}$. Near the ask price, the net volume of
incoming limit orders is five times larger than the volume stored in
the average order book. More than one $\sigma_G$ away from
the bid and ask price, the order flow decreases rapidly.
Integration of the order flow density up to a given return $G$
contributes the additional volume $Q_{\rm flow}(G) = \int_0^{G}
\rho_{\rm flow} (\gamma) \; d\gamma$, which is displayed in
Fig.~\ref{gen_flow.fig}a. It grows fast  for small returns and
saturates for larger returns.

Furthermore, there is an additional volume of incoming limit
orders generated by the returns $G$ due to the anticorrelation
between returns and limit orders. The density of these additional
orders is described by the conditional expectation value
%
%*************************** price feedback *****************************
\begin{equation}
\rho_{\rm c}(\gamma,G)\! =\! \left<Q^{\rm add}_{
t_0}(\gamma)\! -\! Q^{\rm canc}_{t_0}(\gamma)\right>_{G}
 - \left<Q^{\rm add}_{t_0}(\gamma)\! -\!
Q^{\rm canc}_{t_0}(\gamma) \right>   .
\label{feedback.eq}
\end{equation}
%**************************************************************************
%
Here, $Q^{\rm add}_{t_0}(\gamma)$ is the number of limit orders
added to the book at a depth $\gamma$ within the time interval
$[t,t+t_0]$. We find that $\rho_{\rm c}(\gamma,G)$ approximately
saturates for $t_0 \ge 30{\rm min}$.

We consider a situation with "stationary price changes" by assuming
that $G(t)\equiv G$ is constant in time. Then, the choice $t_0= 30
{\rm min}$ makes sure that also the additional limit order volume due
to returns in past time intervals is taken into account. The
correlation volume corresponding to a return $G$ is
%
%************************  correlation volume **************************
\begin{equation}
Q_{\rm corr}(G) = \int_0^G \rho_{\rm c}(\gamma,G) d \gamma \ .
\end{equation}
%***********************************************************************
%
$Q_{\rm corr}(G)$ is slightly negative for small $G$ and increases
then almost linearly for larger $G$ (see Fig.4a).

The total volume $Q(G)$ corresponding to a
return $G$ is the sum
%
%**************************** total volume ***************************
\begin{equation}
Q(G) = Q_{\rm book}(G) + Q_{\rm flow}(G) + Q_{\rm corr}(G)
\label{totvolume.eq}
\end{equation}
%*********************************************************************
%
of the volume $Q_{\rm book}(G)$ of orders stored in the limit order
book up to a depth $G$, the volume $Q_{\rm flow}(G)$ arriving within a
five minute interval on average, and the correlation volume $Q_{\rm
  corr}(G)$.  The theoretical price impact function $I_{\rm
  theory}(Q)$ calculated by inverting this relation is shown in
Fig.~\ref{gen_flow.fig}b.

The agreement between $I_{\rm theory}(Q)$ and $I_{\rm market}(Q)$ is
excellent, up to $G = 10 \sigma_G$ there are no deviations within the
error bars of $I_{\rm market}(Q)$. The additional liquidity due the
influx of limit orders correlated with past returns has a very strong
influence on the price impact of market orders. It strongly reduces
the virtual price impact and is responsible for the
empirically observed concave shape of the price impact function.  We
note that a reduction of ``bare'' price impact by liquidity providers
was recently postulated in \cite{Bou+03} in order to reconcile the
strong autocorrelations of market orders with the uncorrelated random
walk of returns, and that  \cite{LiFa03} explains the uncorrelated nature
of returns by liquidity fluctuations.

In summary, we find that the virtual price impact function as
calculated from the average order book is convex and increases much
faster than the concave price impact function for market orders.  This
difference can be explained by taking into account dynamical
properties of the order book, i.e.~the average net order flow and the
strong anticorrelation between returns and limit order flow. This
anticorrelation leads to an additional influx of limit orders as a
reaction to price changes, which reduces the price impact of market
orders.  Including these dynamical effects, we quantitatively model
the price impact of market orders.

{\it Acknowledgments:} We thank J.-P.~Bouchaud, M.~Potters, and
D.~Stauffer for helpful discussions.


\vspace{-0.5cm}

\begin{thebibliography}{99}

\vspace{-0.5cm}


\bibitem{Takayasu02}  H.~Takayasu, ed., {\em Empirical Analysis of
Financial Fluctuations.} Springer, Berlin, 2002.

\bibitem{Hasbrouck91} J.~Hasbrouck,
%Measuring the Information Content of Stock Trades
J. Finance {\bf 46}, 179-207 (1991).

\bibitem{HaLoMc92} J.~Hausmann, A.~Lo, and C.~MacKinlay, J. Finan. Econ.
 {\bf 31}, 319-379 (1992).

\bibitem{kempf99} A.~Kempf and O.~Korn, J.  Finan. Mark.  \textbf{2},
29-48  (1999).

\bibitem{pler2002} V.~Plerou, P.~Gopikrishnan, X.~Gabaix, and
H.E.~Stanley,
%Quantifying Stock Price Response to Demand Fluctuations
Physical Review E \textbf{66}, 027104[1]-027104[4] (2002).

\bibitem{Rosenow02} B.~Rosenow, Int. J. Mod. Phys. C {\bf 13}, 419-425 (2002).

\bibitem{EvLy02} M.D.~Evans and R.K.~Lyons,
% Order Flow and Exchange Rate Dynamics
J. Political Economy {\bf 110}, 170-180 (2002).

\bibitem{LiFaMa03} F.~Lillo, J.D.~Farmer, and R.N.~Mantegna,
%Master curve for price-impact function
Nature {\bf 421}, 129 (2003).

\bibitem{Ga+03} X.~Gabaix, P.~Gopikrishnan, V.~Plerou, and H.E.~Stanley,
%A theory of power-law distributions in financial market fluctuations
Nature {\bf 423}, 267-270 (2003).


\bibitem{potbou2002} M.~Potters, J.-P.~Bouchaud,
%More statistical properties of order books and price impact
%eprint cond-mat/0210710 (2002).
Physica A {\bf 324}, 133-140 (2003).

\bibitem{Hop02} C.~Hopman, {\em Are supply and demand driving stock prices?},
MIT working paper, Dec.~2002.

\bibitem{Bou+03} J.-P.~Bouchaud, Y.~Gefen, M.~Potters, and M.~Wyart,
%Fluctuations and response in financial markets: the subtle nature of `random'
%price changes
eprint cond-mat/0307332 (2003).

\bibitem{LeRe91} C.M.~Lee and M.J.~Ready,
%Inferring Trade Direction from Intraday Data
J. Finance {\bf 46}, 733-746 (1991).

\bibitem{ticker} We analyzed the following companies (ticker
symbols): AMAT, BRCD, BRCM, CSCO, INTC, KLAC, MSFT, ORCL, QLGC,
SEBL.

\bibitem{hidden}  We do not include market orders executing "hidden"
limit orders in the definition of $Q(t)$ as we want to compare with the
order book containing "visible" orders only.

\bibitem{Zhang99}  Y.-C.~Zhang, Physica A {\bf 269}, 30 (1999).


\bibitem{MaMi01} S.~Maslov and M.~Mills,
%Price fluctuations from the order book perspective - empirical facts
%and a simple model
Physica A {\bf 299}, 234-246 (2001).


\bibitem{ChSt01} D.~Challet and R.~Stinchcombe,
% Analyzing and modelling 1+1d markets, eprint cond-mat/0106114.
Physica A {\bf 300}, 287-299 (2001).

\bibitem{Bouchaud02} J.-P.~Bouchaud, M.~M\'ezard, and M.~Potters,
%Statistical properties of stock order books: empirical results and models
Quantitative Finance, \textbf{2},  251-256 (2002).



\bibitem{average} We find that the function $\langle I _{\rm
    book}\rangle (Q)$ obtained by calculating the price impact at a
  given moment in time and doing the average afterwards is steeper
  than $I_{\rm book} (Q)$.  This difference seems to be due to the
  fact that $\langle I_{\rm book} \rangle(Q)$ is dominated by rare
  events with low liquidity, and that the average is very different
  from the typical function, which is better described by $I_{\rm
    book} (Q)$.


\bibitem{Farmer} M.G.~Daniels, J.D.~Farmer, L.~Gillemot, G.~Iori, and
E.~Smith,
%Quantitative Model of Price Diffusion and Market Friction Based on Trading
%as a Mechanistic Random Process
PRL {\bf 90}, 108102[1]-108102[4] (2003).

\bibitem{LiFa03} F.~Lillo and J.D.~Farmer,
%The long memory of the efficient market
eprint cond-mat/0311053.

\end{thebibliography}
\end{document}